\documentclass[%
 reprint,
superscriptaddress,
 amsmath,amssymb,
 aps,
pra,
]{revtex4-1}

\usepackage{amsthm,amssymb}
\usepackage{amsmath}
\usepackage{graphicx}
\usepackage{dcolumn}
\usepackage{bm}
\usepackage{multirow} 
\usepackage{mathdots}
\usepackage{enumerate}
\usepackage{algorithm}
\usepackage{algorithmic}
\usepackage{color}
\usepackage{epsfig}
\usepackage{subfigure}
\usepackage{caption}
\usepackage{booktabs}

\theoremstyle{remark}

\begin{document}

\makeatletter
\newcommand{\ud}{\mathrm{d}}
\newcommand{\rmnum}[1]{\romannumeral #1}
\newcommand{\Rmnum}[1]{\expandafter\@slowromancap\romannumeral #1@}
\newcommand{\udots}{\mathinner{\mskip1mu\raise1pt\vbox{\kern7pt\hbox{.}}
        \mskip2mu\raise4pt\hbox{.}\mskip2mu\raise7pt\hbox{.}\mskip1mu}}
\makeatother

\preprint{APS/123-QED}

\title{Quantum Algorithm for Unsupervised Anomaly Detection}

\author{Ming-Chao Guo}
\affiliation{State Key Laboratory of Networking and Switching Technology, Beijing University of Posts and Telecommunications, Beijing, 100876, China}
\author{Shi-Jie Pan}
\affiliation{State Key Laboratory of Networking and Switching Technology, Beijing University of Posts and Telecommunications, Beijing, 100876, China}
\author{Wen-Min Li}
\email{liwenmin@bupt.edu.cn}
\author{Fei Gao}
\email{gaof@bupt.edu.cn}
\author{Su-Juan Qin}
\affiliation{State Key Laboratory of Networking and Switching Technology, Beijing University of Posts and Telecommunications, Beijing, 100876, China}
\author{Xiao-Ling Yu}
\affiliation{Comprehensive Research Center of Electronic Information Technology in the MIIT, Shandong, 264200, China}
\author{Xuan-Wen Zhang}
\affiliation{Comprehensive Research Center of Electronic Information Technology in the MIIT, Shandong, 264200, China}
\author{Qiao-Yan Wen}
\affiliation{State Key Laboratory of Networking and Switching Technology, Beijing University of Posts and Telecommunications, Beijing, 100876, China}

\date{\today}

\begin{abstract}
Anomaly detection, an important branch of machine learning, plays a critical role in fraud detection, health care, intrusion detection, military surveillance, etc. As one of the most commonly used unsupervised anomaly detection algorithms, the Local Outlier Factor algorithm (LOF algorithm) has been extensively studied. This algorithm contains three steps, i.e., determining the k-distance neighborhood for each data point $\bm{x}$, computing the local reachability density of $\bm{x}$, and calculating the local outlier factor of $\bm{x}$ to judge whether $\bm{x}$ is abnormal. The LOF algorithm is computationally expensive when processing big data sets. Here we present a quantum LOF algorithm consisting of three parts corresponding to the classical algorithm. Specifically, the k-distance neighborhood of $\bm{x}$ is determined by amplitude estimation and minimum search; the local reachability density of each data point is calculated in parallel based on the quantum multiply-adder; the local outlier factor of each data point is obtained in parallel using amplitude estimation. It is shown that our quantum algorithm achieves exponential speedup on the dimension of the data points and polynomial speedup on the number of data points compared to its classical counterpart. This work demonstrates the advantage of quantum computing in unsupervised anomaly detection.
\end{abstract}
\pacs{Valid PACS appear here}
\maketitle

\section{Introduction}
\label{}
Quantum computing theoretically demonstrates its computational advantages in solving certain problems
compared with classical computing, such as factoring integers \cite{P1994}, searching in unstructured databases \cite{L1996}, solving linear and differential equations \cite{AAS2009,LCS2018,HYL2021}, attacking cryptography \cite{ZBH2022,HCB2022,BFG2022}, and designing cryptographic protocols for the private query \cite{CXT2020,FSW2019,VSL2008}. Recently, the combination of quantum computing and machine learning, Quantum Machine Learning (QML) \cite{JPN2017,AJS2019}, has emerged as a promising application of quantum technology. QML made great strides in data classification \cite{SMP2013,NDS2012}, clustering \cite{SBS2017}, 
linear regression \cite{G2017,CFQ2019,CFC2019}, association rule mining \cite{CFQ2016}, and dimensionality reduction \cite{SMP2014,SLH2020,CFS2019}.

Anomaly Detection (AD) refers to finding patterns in data that do not conform to expected behavior. It has been extensively used in various fields, such as credit card fraud detection \cite{EB1997}, intrusion detection \cite{VP2005}, and health care \cite{SL2001}. Over the years, many AD algorithms have been proposed, which can be divided into supervised, semi-supervised, and unsupervised \cite{VAV2009}. Among them, unsupervised AD algorithms are more widely applicable as they do not require labeled data \cite{AMS2011}. Breunig et al. proposed a density-based anomaly detection algorithm (called LOF algorithm) \cite{MHR2000}, one of the most widely used unsupervised AD algorithms. The advantage of this algorithm is that it performs well when dealing with datasets containing patterns with diverse characteristics. However, similar to other AD algorithms, the LOF algorithm is quite time-consuming when processing big data sets.

In recent years, some researchers successfully combined AD algorithms with quantum techniques and obtained various degrees of speedups. In 2018, Liu et al. proposed a quantum kernel principal component analysis algorithm \cite{NP2018}, which achieves an exponential speedup on the dimension of the training data compared to the classical counterpart. In 2019, Liang et al. presented a quantum anomaly detection with density estimation algorithm \cite{JSM2019}. Its complexity is logarithmic in the dimension and the number of training data compared to the corresponding classical algorithm. In 2022, Guo et al. proposed a quantum algorithm for anomaly detection \cite{MHY2022}, which achieves exponential speedup on the number of training data points over its classical counterpart. 

In this paper, given the importance of unsupervised anomaly detection, we focus on the research of quantum algorithms for unsupervised anomaly detection. Specifically, we propose a quantum LOF algorithm. The classical LOF algorithm contains three steps: (1) determines the k-distance neighborhood of each data point; (2) computes the reachable distance between each data point and its k-distance neighbors and obtains the local reachability density by computing the mean of these reachability distances; (3) obtains the local outlier factor of each data point by comparing its local reachability density with its neighbors. Our quantum algorithm also consists of three parts, corresponding to the three steps of the classical LOF algorithm. In the first one, amplitude estimation \cite{GPM2002} and minimum search \cite{DP1996} are utilized to speed up the process of determining the k-distance neighborhood. Given the information of the k-distance neighborhood, the second one accesses this information using QRAM \cite{GSL2008,GVS2008} and then calculates the local reachability density of each data point in parallel based on the quantum multiply-adder \cite{LJ2017,STJ2017}. The third one executes the quantum amplitude estimation to obtain local outlier factors in parallel and performs Grover's algorithm \cite{L1996} to speed up the search for abnormal data points. As a conclusion, our quantum algorithm can achieve exponential speedup on the dimension of the data points $n$ and polynomial speedup on the number of data points $m$ compared to its classical counterpart.

The rest of this article is organized as follows. In Sec.~\ref{sec2}, we briefly review the classical LOF algorithm. In Sec.~\ref{sec3}, we present a quantum LOF algorithm and analyze its complexity in detail. The conclusion is given in Sec.~\ref{sec4}.

\section{Review of the LOF algorithm}
\label{sec2}
In this section, we introduce the relevant definitions and the classical LOF algorithm \cite{MHR2000}.

\subsection{Definitions}
\label{sec2.1}
For convenience, we begin with the following related definitions \cite{MHR2000}:

\textbf{Definition 1}: k-distance: For any positive integer $k$, the k-distance of $\bm{x}\in D$ ($D$ is an unlabeled dataset), denoted as $k_{-}d(\bm{x})$, is defined as the distance $d(\bm{x,y})$ between $\bm{x}$ and $\bm{y}\in D$ such that: there are at least $k$ data points $\bm{y'}\in D/\{\bm{x}\}$ that meet $d(\bm{x,y'})\le d(\bm{x,y})$ and at most $k-1$ data points $\bm{y'}\in D/\{\bm{x}\}$ that meet $d(\bm{x,y'})<d(\bm{x,y})$.

\textbf{Definition 2}: k-distance neighborhood: Given the $k_{-}d(\bm{x})$, the k-distance neighborhood of $\bm{x}$ is expressed as $N_{k}(\bm{x})=\{\bm{y}\in D/{\bm{x}}~|~d(\bm{x,y})\le k_{-}d(\bm{x})\}$. These data points $\bm{y}$ are called the k-distance neighbors of $\bm{x}$.

\textbf{Definition 3}: reachability distance: Given the parameter $k$, the reachability distance of data point $\bm{x}$ with respect to $\bm{y}$ is defined as $reach\mbox{-}dist_{k}(\bm{x,y})=\max\{k\mbox{-}distance(\bm{y}),d(\bm{x,y})\}$.

Intuitively, if $\bm{x}$ is far away from $\bm{y}$, then the reachability distance between the two is simply their actual distance. However, if they are ``sufficiently'' close, the actual distance is replaced by the k-distance of $\bm{y}$. The purpose of introducing reachability distance is to reduce statistical fluctuation in the distance measure.

\textbf{Definition 4}: local reachability density: The local reachability density of $\bm{x}$ is defined as
\begin{equation}
lrd_{k}(\bm{x})=\bigg(\frac{\sum_{\bm{y}\in N_{k}(\bm{x})}reach\mbox{-}dist_{k}(\bm{x,y})}{|N_{k}(\bm{x})|}\bigg)^{-1}.
\end{equation}
Intuitively, the local reachability density of a data point $\bm{x}$ is the inverse of the average reachability distance based on the k-distance neighbors of $\bm{x}$.

\textbf{Definition 5}: local outlier factor: The local outlier factor of $\bm{x}$ is defined as
\begin{equation}
LOF_k(\bm{x})=\frac{1}{|N_{k}(\bm{x})|}\sum_{\bm{y}\in N_{k}(\bm{x})}\frac{lrd_{k}(\bm{y})}{lrd_{k}(\bm{x})}.
\end{equation}
The local outlier factor (LOF) of $\bm{x}$ captures the degree to which we call $\bm{x}$ an outlier. It is the average ratio of the local reachability density of $\bm{x}$ and those of $\bm{x}$'s k-distance neighbors. It is easy to see that the lower $\bm{x}$'s local reachability density, the higher the local reachability densities of $\bm{x}$'s k-distance neighbors are, and the higher the LOF value of $\bm{x}$.

\subsection{LOF algorithm}
\label{sec2.1}
Given an unlabeled data set $X=\{\bm{x^i}\}_{i=1}^m$ and parameter $k$, where $\bm{x^i}=(x^i_1,x^i_2,\cdots,x^i_n)\in R^n$. The LOF algorithm contains three steps: the first step is to find the k-distance neighborhood of each data point; the second step is to calculate the reachable distance between each data point and its k-distance neighbors, and then obtain the local reachable density by computing the mean of these reachable distances; the third step is to get the local outlier factor of each data point by comparing its local reachable density with the data point in its neighborhood. We identify whether each data is an anomaly by comparing its local outlier factor with a pre-determined threshold $\delta$. The whole procedure of LOF algorithm is shown in Algorithm 1.
\begin{table}[htbp]
\begin{center}
\begin{tabular*}{\hsize}{@{}@{\extracolsep{\fill}}l@{}}
\toprule[1.5pt]
\textbf{Algorithm 1} ~The procedure of LOF algorithm\\
\midrule[0.3pt]
\hline
\quad \textbf{Input:}~The data set $X$ and threshold $\delta$;\\
\\
\quad (1) Determine the k-distance neighborhood $N_{k}(\bm{x^i})$ of the \\
\quad~~~~~data point $\bm{x^i};$\\
\quad (2) Calculate the local reachability density $lrd_{k}(\bm{x^i})$ of the \\
\quad ~~~~~data point $\bm{x^i}$ by Eq. (1);\\
\quad (3) Compute the local outlier factor $LOF_k(\bm{x^i})$ of the \\
\quad ~~~~~data point $\bm{x^i}$ by Eq. (2);\\
\quad If $LOF_k(\bm{x^i})\geq\delta$, the data point $\bm{x^i}$ is marked as an anomaly.\\
\\
\quad \textbf{Output:}~Abnormal data points. \\
\hline
\bottomrule[1.5pt]
\end{tabular*}
\label{Table1}
\end{center}
\end{table}

The complexity of step (1) is $O(m^{2}nk)$, the complexity of step (2) and step (3) is $O(mk)$. In other words, the LOF algorithm is computationally expensive when processing big data sets.

\section{Quantum LOF algorithm}
\label{sec3}
In this section, we propose a quantum version of the LOF algorithm and analyze its complexity. Our algorithm consisting of three parts corresponding to the classical LOF algorithm. Firstly, we utilize amplitude estimation \cite{GPM2002} and minimum search \cite{DP1996} to speed up the process of finding the k-distance neighborhood. Secondly, we use a data structure of QRAM \cite{GSL2008} to access the information of k-distance neighborhood and then calculate the local reachability density of each data point in parallel based on the quantum multiply-adder \cite{LJ2017,STJ2017}. Finally, we can obtain local outlier factors in parallel through the quantum amplitude estimation and perform Grover's algorithm \cite{L1996} to search for abnormal data points satisfying $LOF_k(\bm{x^i})\geq\delta$ in Sec.~\ref{sec3.3}. An overview of our algorithm is shown in Algorithm 2.
\begin{table*}[htbp]
\begin{center}
\begin{tabular*}{\hsize}{@{}@{\extracolsep{\fill}}l@{}}
\toprule[1.5pt]
\textbf{Algorithm 2}~ The procedure of quantum LOF algorithm\\
\midrule[1pt]
\quad \textbf{Input:}~~Data matrix $X$ stored in a QRAM. Pre-determined threshold $\delta$ and parameter $k$;\\
\\
\quad Step 1: Determine the k-distance neighborhood $N_{k}(\bm{x^i})$ of the data point $\bm{x^i}$ to get \\
\quad ~~~~~~~~~~~~~~~~~~~~~~~~~~~~~$|i\rangle\frac{1}{\sqrt{n_i}}\sum_{\bm{x^t}\in N_{k}(\bm{x^i})}|t\rangle|\frac{1}{\sqrt{n}C}d(\bm{x^i},\bm{x^t})\rangle|k_{-}d(\bm{x^i})\rangle|n_i\rangle$.\\
\quad ~~~~~~~~~~~Measure the quantum state to obtain the information of $N_{k}(\bm{x^i})$;\\
\quad Step 2: Calculate the local reachability density $lrd_{k}(\bm{x^i})$ of the data point $\bm{x^i}$ to get\\
\quad ~~~~~~~~~~~~~~~~~~~~~~~~~~~~~$\frac{1}{\sqrt{m}}\sum_{i=1}^m|i\rangle|\big[lrd_{k}(\bm{x^i})\big]^{-1}\rangle\frac{1}{\sqrt{n_i}}\sum_{j=1,\bm{x^t}\in
N_{k}(\bm{x^i})}^{n_i}|j\rangle|t\rangle|\big[lrd_{k}(\bm{x^t})\big]^{-1}\rangle$;\\
\quad Step 3: Compute the local outlier factor $LOF_k(\bm{x^i})$ of data point $\bm{x^i}$ to obtain\\
\quad ~~~~~~~~~~~~~~~~~~~~~~~~~~~~~$\frac{1}{\sqrt{m}}\sum_{i=1}^m|i\rangle|LOF_{k}(\bm{x^i})\rangle$;\\
\quad Grover's algorithm is applied to search all indices $i$ of abnormal data points that satisfy $LOF_k(\bm{x^i})\geq\delta$. \\
\\
\quad \textbf{Output:}~Abnormal data points. \\
\bottomrule[1.5pt]
\end{tabular*}
\label{Table2}
\end{center}
\end{table*}

Assume that the data set $X=\{\bm{x^i}\}_{i=1}^m$ is stored in a QRAM \cite{GSL2008,GVS2008}, which allows the following mappings to be performed in time $O(\log mn)$:
\begin{equation}
O_X:|i\rangle|j\rangle|0\rangle\rightarrow|i\rangle|j\rangle|x^i_j\rangle,
\end{equation}
where $i=1,2,\cdots,m,j=1,2,\cdots,n$.

As a common tool, QRAM has been used to handle state preparation tasks in most of the quantum algorithms, espectially the quantum machine learning algorithms, such as data classification \cite{SMP2013,NDS2012}, clustering \cite{SBS2017}, linear regression \cite{G2017,CFQ2019,CFC2019}, association rule mining \cite{CFQ2016}, dimensionality reduction \cite{SMP2014,SLH2020,CFS2019}.

\textbf{Definition 6:} Quantum adder \cite{STJ2017,DT2000}: Let $x_1x_2\cdots x_n$ be the binary representation for $x$, where $x=x_1\cdot2^{n-1}+x_2\cdot2^{n-2}+\cdots+x_n\cdot2^0$, $|x\rangle=|x_1\rangle|x_2\rangle\cdots|x_n\rangle$, and $|y\rangle=|y_1\rangle|y_2\rangle\cdots|y_n\rangle$. The quantum adder can realize the following transformation,
\begin{equation}
|x\rangle|y\rangle|0\rangle\xrightarrow{quantum~adder}|x\rangle|y\rangle|x+y\rangle.
\end{equation}
The circuit diagram of the quantum adder is shown in Refs. \cite{STJ2017,DT2000}. It consumes $O(n^2)$ controlled rotation gates, so the complexity of the quantum adder is $O(n^2)$.

\textbf{Definition 7:} Quantum multiply-adder \cite{STJ2017}: Let $|x\rangle=|x_1\rangle|x_2\rangle\cdots|x_n\rangle$, $|y\rangle=|y_1\rangle|y_2\rangle\cdots|y_n\rangle$ and $|z\rangle=|z_1\rangle|z_2\rangle\cdots|z_{2n}\rangle$, the quantum multiply-adder can
realize the following transformation,
\begin{equation}
|x\rangle|y\rangle|z\rangle\xrightarrow{quantum~multiply\mbox{-}adder}|x\rangle|y\rangle|z+x\cdot y\rangle.
\end{equation}
Its circuit diagram is shown in Ref. \cite{STJ2017}. The complexity of the quantum multiply-adder is $O(n^3)$.
\subsection{ Quantum process of finding the k-distance neighborhood}
\label{sec3.1}

The k-distance neighborhood of each data point is determined based on its k-distance (as shown in Definition 1 and Definition 2). 
The choice of parameter $k$ is crucial to the performance of such algorithms, but how to choose a suitable $k$ is outside the scope of our discussion. Here we assume that the parameter $k$ is given in advance.

\subsubsection{Step details}

We adopt the quantum amplitude estimation and quantum minimum search to determine the k-distance neighborhood of $\bm{x^i}$. The whole procedure is depicted as follows.

(1.1). Prepare the quantum state ($i=1,2,\cdots,m$)
\begin{equation}
|i\rangle\frac{1}{\sqrt{(m-1)n}}\sum_{t\neq i}\sum_{j=1}^n|t\rangle|j\rangle|0\rangle|0\rangle.
\end{equation}

(1.2) Apply the oracle $O_X$ to prepare
\begin{equation}
|i\rangle\frac{1}{\sqrt{(m-1)n}}\sum_{t\neq i}\sum_{j=1}^n|t\rangle|j\rangle|x_{j}^i\rangle|x_{j}^t\rangle.
\end{equation}

(1.3) Perform the the quantum multiply-adder (QMA) \cite{LJ2017,STJ2017} to creat
\begin{equation}
|i\rangle\frac{1}{\sqrt{(m-1)n}}\sum_{t\neq i}\sum_{j=1}^n|t\rangle|j\rangle|x_j^i\rangle|x^t_j\rangle|x_{j}^i-x_{j}^t\rangle.
\end{equation}

(1.4) Append an ancillary qubit and perform controlled rotation from $|0\rangle$ to $\frac{x_j^i-x_j^t}{C}|0\rangle+\sqrt{1-(\frac{x_j^i-x_j^t}{C})^2}|1\rangle$ conditioned on $|x_{j}^i-x_{j}^t\rangle$ \cite{AAS2009}. Uncompute the fourth, fifth and sixth registers to generate
\begin{align}
&|i\rangle\frac{1}{\sqrt{(m-1)n}}\sum_{t\neq i}\sum_{j=1}^n|t\rangle|j\rangle(\frac{x_j^i-x_j^t}{C}|0\rangle+\sqrt{1-(\frac{x_j^i-x_j^t}{C})^2}|1\rangle)\nonumber\\
&:=|i\rangle\frac{1}{\sqrt{(m-1)}}\sum_{t\neq i}|t\rangle|\chi_t\rangle,
\end{align}
where $C=\max|x_j^i-x_j^t|$. The state $|\chi_t\rangle$ can be rewritten as $|\chi_t\rangle=\sin\theta_t|\chi_t^0\rangle+\cos\theta_t|\chi_t^0\rangle$, where $\sin\theta_t|\chi_t^0\rangle=\frac{1}{\sqrt{n}}\sum_{j=1}^d\frac{x^i_j-x^t_j}{C}|j\rangle|0\rangle$, $\cos\theta_t|\chi_t^1\rangle=\frac{1}{\sqrt{n}}\sum_{j=1}^n\sqrt{1-(\frac{x^i_j-x^t_j}{C})^2}|j\rangle|1\rangle$. Thus, we have $\sin^2\theta_t=\frac{1}{n}\sum_{j=1}^n(\frac{x^i_j-x^t_j}{C})^2, \theta_t\in[0,\frac{\pi}{2}]$.

(1.5) Perform the quantum amplitude estimation to get
\begin{equation}
|i\rangle\frac{1}{\sqrt{(m-1)}}\sum_{t\neq i}|t\rangle|\chi_t\rangle|\frac{\theta_t}{\pi}\rangle,
\end{equation}
then perform the sine gate on $|\frac{\theta_t}{\pi}\rangle$ \cite{STJ2017} and uncompute the redundant registers $|\chi_t\rangle$ to create the superposition of the distances between all points and $\bm{x^i}$:
\begin{equation}
|i\rangle\frac{1}{\sqrt{(m-1)}}\sum_{t\neq i}|t\rangle|\sin\theta_t\rangle=|i\rangle\frac{1}{\sqrt{(m-1)}}\sum_{t\neq i}|t\rangle|\frac{1}{\sqrt{n}C}d(\bm{x^i},\bm{x^t})\rangle,
\end{equation}
where $d(\bm{x^i},\bm{x^t})$ represents the Euclidean distance between $\bm{x^i}$ and $\bm{x^t}$. (The detailed procedure can be seen in ref. \cite{MHY2022}.)

(1.6) Append an ancilla register, perform the quantum minimum search \cite{DP1996} to get the k-distance of $\bm{x^i}$ and stored it in the ancilla register:
\begin{equation}
|i\rangle\frac{1}{\sqrt{m-1}}\sum_{t\neq i}|t\rangle|\frac{1}{\sqrt{n}C}d(\bm{x^i},\bm{x^t})\rangle|k_{-}d(\bm{x^i})\rangle.
\end{equation}

(1.7) Perform Grover's algorithm \cite{L1996} to create the superposition state of labels for points in the k-distance neighborhood of $\bm{x^i}$:
\begin{equation}
|i\rangle\frac{1}{\sqrt{n_i}}\sum_{\bm{x^t}\in N_{k}(\bm{x^i})}|t\rangle|\frac{1}{\sqrt{n}C}d(\bm{x^i},\bm{x^t})\rangle|k_{-}d(\bm{x^i})\rangle,
\end{equation}
where $\frac{1}{\sqrt{n}C}d(\bm{x^i},\bm{x^t})\leq k_{-}d(\bm{x^i})$.

(1.8) Perform quantum counting \cite{GPM2002} to get the number of points in the k-distance neighborhood of $\bm{x^i}$:
\begin{equation}
|i\rangle\frac{1}{\sqrt{n_i}}\sum_{\bm{x^t}\in N_{k}(\bm{x^i})}|t\rangle|\frac{1}{\sqrt{n}C}d(\bm{x^i},\bm{x^t})\rangle|k_{-}d(\bm{x^i})\rangle|n_i\rangle,
\end{equation}
where $n_{i}=|N_{k}(\bm{x^i})|$. Then by measuring the state in computational basis for several times, we could obtain the index $t$, the distance $\frac{1}{\sqrt{n}C}d(\bm{x^i},\bm{x^t})$, and the number $n_{i}$ of k-distance neighbors of $\bm{x^i}$ for $i=1,2,\cdots,m$.
\begin{figure*}[htb]
	\centering
		\includegraphics[height=6cm,width=12cm]{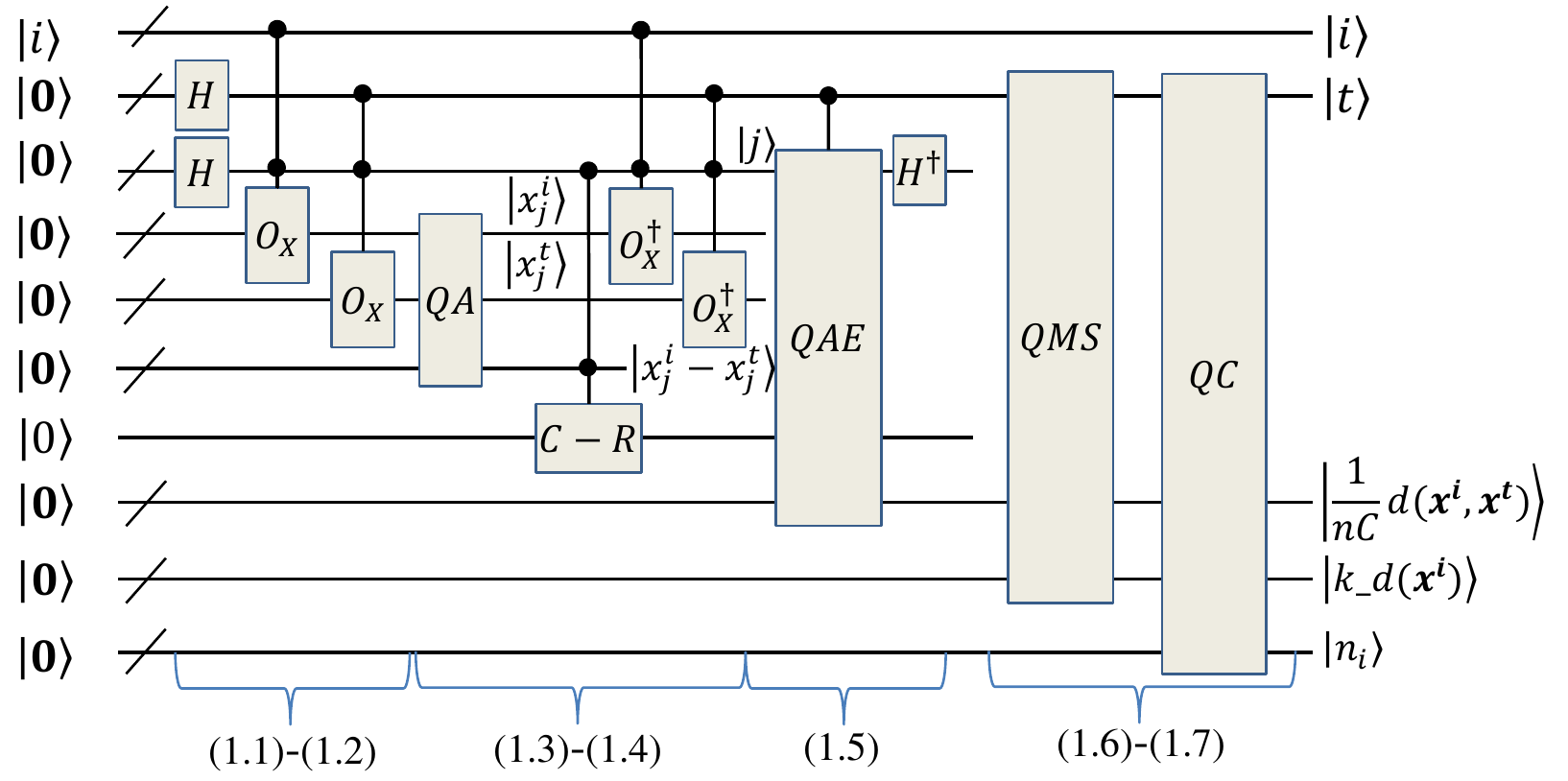}
\caption{Quantum circuit of step 1. Here ``/" denotes a bundle of wires, $QA$ represents a quantum adder, $C-R$ denotes controlled rotation, $QMS$ represents quantum minimum search and $QC$ denotes quantum counting.}
	\label{fig:kappa}
\end{figure*}
\subsubsection{Complexity analysis}

In step (1.2), the oracle $O_X$ can be implemented in $O(\log mn)$. In step (1.3)-(1.4), it takes QMA and control rotation with complexity $O(1)$ (we assume that the data points are represented by a constant number of qubits), the complexity of these gates can be omitted \cite{STJ2017}. In step (1.5), to ensure that the error is $\varepsilon_1$, the amplitude estimation costs $O[(\log mn)/\varepsilon_1]$ time. (For detailed complexity analysis, see the second paragraph of Sec. 3.3 in Ref.~\cite{MHY2022}.) In step (1.6), the quantum minimum search is performed to find the k-distance of $\bm{x^i}$, which requires repeating $O(k\sqrt{m})$ operations of steps (1.2)-(1.5), so the complexity of step (1.6) is $O[k\sqrt{m}\log(mn)/\varepsilon_1]$. In step (1.7), the complexity of performing Grover's algorithm is $O[n_i\sqrt{m}\log(mn)/\varepsilon_1$]. As for the quantum counting in step (1.8), let $|n_i-\hat{n}_i|\le\varepsilon_2$, it needs needs to repeatedly perform $O(\sqrt{(m-1)n_i}/\varepsilon_2)$ operations of step (1.2)-(1.5) and one operation of step (1.6), so the complexity of the quantum counting is $O[\sqrt{mn_i}\log(mn)/\varepsilon_1\varepsilon_2]+O[k\sqrt{m}\log(mn)/\varepsilon_1]$. We can get that the complexity of step (1.8) is $O[n_i\sqrt{m}\log(mn)/\varepsilon_1\varepsilon_2]$. 

Now, we analyze the error of step (1.5). For convenience, we use $\hat{a}$ to denote the estimation of $a$ in the following sections. Let $|\theta_t-\hat{\theta}_t|\le\varepsilon_1$, we can obtain
\begin{align}
&|\frac{1}{\sqrt{n}C}\hat{d}(\bm{x^i},\bm{x^t})-\frac{1}{\sqrt{n}C}d(\bm{x^i},\bm{x^t})|=|\sin\theta_t-\sin\hat{\theta}_t|\nonumber\\
&\le|\theta_t-\hat{\theta}_t|\le\varepsilon_1.
\end{align}
For convenience, we use $\overline{d}(\bm{x^i},\bm{x^t})$  to denote $\frac{1}{\sqrt{n}C}d(\bm{x^i},\bm{x^t})$ in the following sections.

The operations from step (1.1) to step (1.7) need to be repeated $m$ times to obtain the k-distance neighborhood of all data points, so the overall complexity is $O[\max_{i}\{n_i\}\cdot m^{\frac{3}{2}}\log(mn)/\varepsilon_1]$. The complexity of steps (1.1) to (1.7) is illustrated in Table I.
\begin{table}[htbp]
\centering
\label{TAB:table1}
\caption{The time complexity of steps 1 to 7}
\setlength{\tabcolsep}{8mm}{
\begin{tabular}{ccccccc|cc}
\hline
Steps\par&Complexity\\ \hline
(1.1)-(1.2)&$O(\log mn)$\\
(1.3)-(1.4)&$O(1)$\\
(1.5)&$O[\frac{(\log mn)}{\varepsilon_1}]$\\
(1.6)&$O[\frac{k\sqrt{m}\log(mn)}{\varepsilon_1}]$\\
(1.7)&$O[\frac{n_i\sqrt{m}\log(mn)}{\varepsilon_1}]$\\
(1.8)&$O[\frac{n_i\sqrt{m}\log(mn)}{\varepsilon_1\varepsilon_2}]$\\
Total complexity&$O[\frac{\max_{i}\{n_i\}\cdot m^{\frac{3}{2}}\log(mn)}{\varepsilon_1\varepsilon_2}]$\\
 \hline\hline
\end{tabular}}
\label{Table1}
\end{table}

\subsection{Quantum process of computing the local reachability density}
\label{sec3.2}

We have obtained the index, distance and number of neighbors of all data points in the previous process. This information can be represented as a matrix and stored in QRAM. To facilitate quantum access in the subsequent sections, there are allows the following mappings
\begin{align}
&U:|i\rangle|0\rangle|0\rangle\rightarrow|i\rangle|n_i\rangle|k_{-}d(\bm{x^i})\rangle,\notag\nonumber\\
&V:|i\rangle|t\rangle|0\rangle\rightarrow|i\rangle|t\rangle|\overline{d}(\bm{x^i,x^t})\rangle,\nonumber\\
&G:|i\rangle|0\rangle\rightarrow|i\rangle\frac{1}{\sqrt{n_i}}\sum_{j=1}^{n_i}|j\rangle,\nonumber\\
&W:|i\rangle\frac{1}{\sqrt{n_i}}\sum_{j=1}^{n_i}|j\rangle|0\rangle\rightarrow|i\rangle\frac{1}{\sqrt{n_i}}\sum_{j=1,\bm{x^t}\in N_{k}}^{n_i}|j\rangle|t\rangle
\end{align}
in complexity $O$[poly$\log(m\max\{n_i\})]$.

\subsubsection{Step details}
We use QRAM to access the information of the k-distance neighborhood of each point, and then compute the local reachability density of each data point in parallel based on quantum multiply-adder. The whole procedure is depicted as follows.

(2.1) Prepare the following quantum state
\begin{equation}
\frac{1}{\sqrt{m}}\sum_{i=1}^m|i\rangle|0\rangle\bigotimes_{j=1}^{\max\{n_i\}}[|j\rangle|0\rangle].
\end{equation}

(2.2) Perform the oracles $U$ and $W$ to prepare the k-distance of $\bm{x^i}$ and  the number of k-distance neighbors of $\bm{x^i}$, uncompute the redundant registers to create the state
\begin{equation}
\frac{1}{\sqrt{m}}\sum_{i=1}^m|i\rangle|n_i\rangle\bigotimes_{\bm{x^t}\in N_{k}(\bm{x^i})}[|t\rangle|k_{-}d(\bm{x^t})\rangle|0\rangle].
\end{equation}

(2.3) Apply the oracle $V$ to create the distance of $\bm{x^i}$ and $\bm{x^t}$ and store it in the register:
\begin{equation}
\frac{1}{\sqrt{m}}\sum_{i=1}^m|i\rangle|n_i\rangle\bigotimes_{\bm{x^t}\in N_{k}(\bm{x^i})}[|t\rangle|k_{-}d(\bm{x^t})\rangle|\overline{d}(\bm{x^i},\bm{x^t})\rangle].
\end{equation}

(2.4) Append ancillary registers and perform the controlled operation $U_f$ on $\bigotimes_{\bm{x^t}\in N_{k}(\bm{x^i})}|k_{-}d(\bm{x^t})\rangle|\overline{d}(\bm{x^i},\bm{x^t})\rangle$, where $U_f:|a\rangle|b\rangle|0\rangle\rightarrow|a\rangle|b\rangle|\max\{a,b\}\rangle$ and the function $f(a,b)=\max\{a,b\}$, to generate the state
\begin{align}
&\frac{1}{\sqrt{m}}\sum_{i=1}^m|i\rangle|n_i\rangle\bigotimes_{\bm{x^t}\in N_{k}(\bm{x^i})}[|t\rangle|k_{-}d(\bm{x^t})\rangle|\overline{d}(\bm{x^i},\bm{x^t})\rangle\nonumber\\
&|\overline{reach\mbox{-}dist}_{k}(\bm{x^i},\bm{x^t})\rangle],
\end{align}
where $\overline{reach\mbox{-}dist}_{k}(\bm{x^i},\bm{x^t})=\max\{k_{-}d(\bm{x^t}),\overline{d}(\bm{x^i,x^t})\}=\frac{1}{\sqrt{n}C}\cdot reach\mbox{-}dist_{k}(\bm{x^i},\bm{x^t})$.

(2.5) Perform the quantum multiply-adder (QMA) on $\bigotimes_{\bm{x^t}\in N_{k}(\bm{x^i})}|\overline{reach\mbox{-}dist}_{k}(\bm{x^i},\bm{x^t})\rangle$ and uncompute redundant registers to create
\begin{align}
&\frac{1}{\sqrt{m}}\sum_{i=1}^m|i\rangle|n_i\rangle|\frac{\sum_{\bm{x^t}\in N_{k}(\bm{x^i})}\overline{reach\mbox{-}dist}_{k}(\bm{x^i},\bm{x^t})}{n_i}\rangle\nonumber\\
&=\frac{1}{\sqrt{m}}\sum_{i=1}^m|i\rangle|n_i\rangle|\big[\overline{lrd}_{k}(\bm{x^i})\big]^{-1}\rangle,
\end{align}
where $\overline{lrd}_{k}(\bm{x^i})=\frac{1}{\sqrt{n}C}lrd_{k}(\bm{x^i})$. 

(2.6) Append registers and perform the oracle $G$ to generate
\begin{equation}
\frac{1}{\sqrt{m}}\sum_{i=1}^m|i\rangle|n_i\rangle|\big[\overline{lrd}_{k}(\bm{x^i})\big]^{-1}\rangle\frac{1}{\sqrt{n_i}}\sum_{j=1}^{n_i}|j\rangle.
\end{equation}

(2.7) Append an ancilla register and execute the oracle $W$ on Eq. (22) to get
\begin{equation}
\frac{1}{\sqrt{m}}\sum_{i=1}^m|i\rangle|n_i\rangle|[\big[\overline{lrd}_{k}(\bm{x^i})\big]^{-1}\rangle\frac{1}{\sqrt{n_i}}\sum_{j=1,\bm{x^t}\in N_{k}(\bm{x^i})}^{n_i}|j\rangle|t\rangle.
\end{equation}

(2.8) Perform operations similar to steps (2.1)-(2.3) on $|t\rangle$ to prepare the following quantum state
\begin{align}
&\frac{1}{\sqrt{m}}\sum_{i=1}^m|i\rangle|n_i\rangle|\big[\overline{lrd}_{k}(\bm{x^i})\big]^{-1}\rangle\frac{1}{\sqrt{n_i}}\sum_{j=1,\bm{x^t}\in N_{k}(\bm{x^i})}^{n_i}|j\rangle|t\rangle|n_t\rangle\nonumber\\
&\bigotimes_{\bm{x^l}\in N_{k}(\bm{x^t})}[|l\rangle|k_{-}d(\bm{x^l})\rangle|\overline{d}(\bm{x^l},\bm{x^t})\rangle|0\rangle].
\end{align}

(2.9) Repeat the operations of step (2.4)-(2.5) to obtain the local density of data $\bm{x^t}$, we can create the state
\begin{equation}
\frac{1}{\sqrt{m}}\sum_{i=1}^m|i\rangle|\big[\overline{lrd}_{k}(\bm{x^i})\big]^{-1}\rangle\frac{1}{\sqrt{n_i}}\sum_{j=1,\bm{x^t}\in N_{k}(\bm{x^i})}^{n_i}|j\rangle|t\rangle|\big[\overline{lrd}_{k}(\bm{x^t})\big]^{-1}\rangle.
\end{equation}
The entire quantum circuit is shown in Fig. 2.
\begin{figure*}[htb]
	\centering
		\includegraphics[height=8cm,width=13cm]{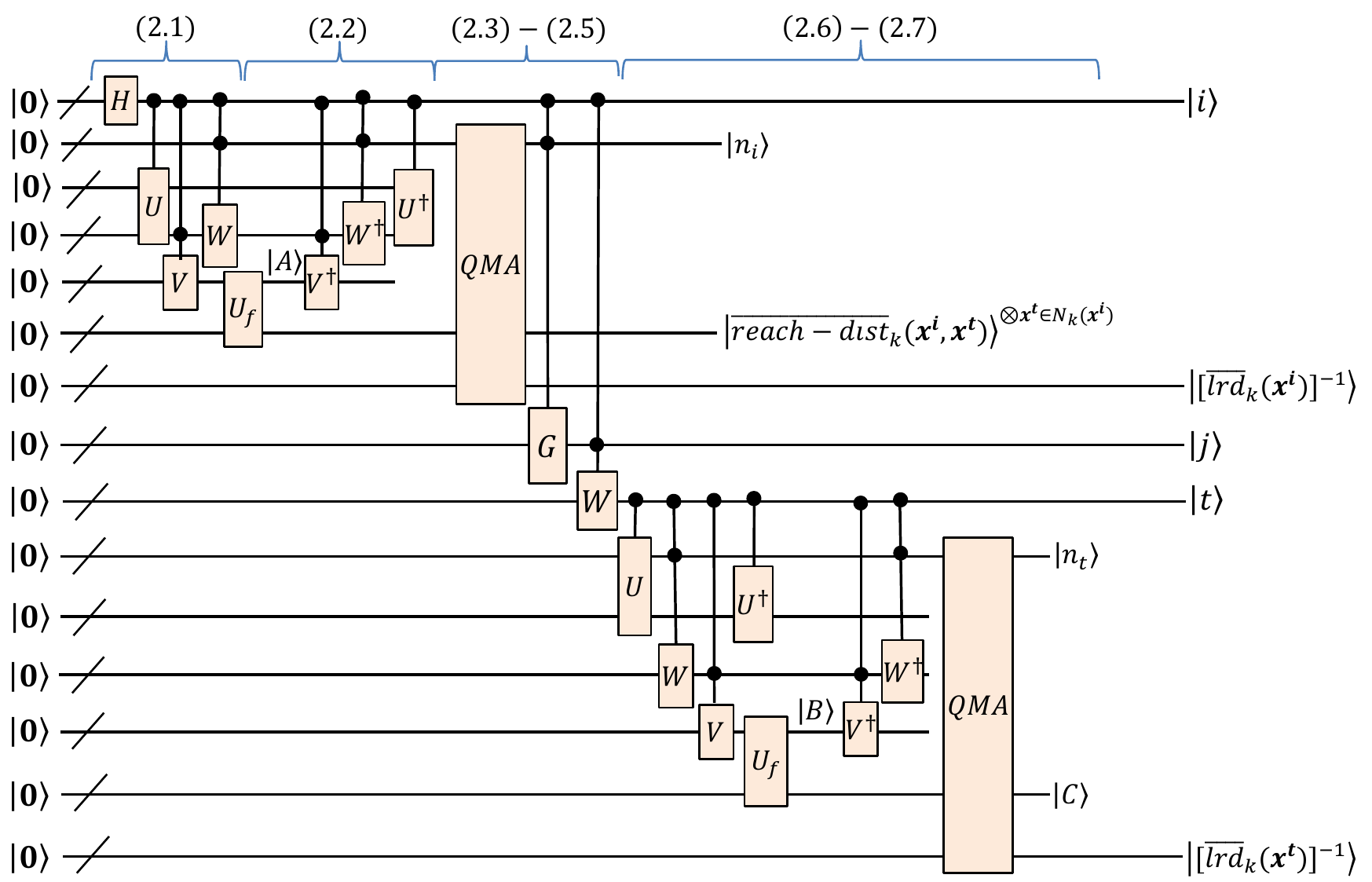}
\caption{Quantum circuit of step 2. Here ``/" denotes a bundle of wires, $QMA$ represents a quantum multiply-adder, $|A\rangle=|\overline{d}(\bm{x^i},\bm{x^t})\rangle^{\otimes\bm{x^t}\in N_k(\bm{x^i})}$, $|B\rangle=|\overline{d}(\bm{x^t},\bm{x^l})\rangle^{\otimes\bm{x^i}\in N_k(\bm{x^t})}$ and $|C\rangle=|\overline{reach\mbox{-}dist}_{k}(\bm{x^i},\bm{x^t})\rangle^{\otimes\bm{x^l}\in N_k(\bm{x^t})}$.}
	\label{fig:kappa}
\end{figure*}
\subsubsection{Complexity analysis}
In step (2.1), to prepare the state in Eq. (17), $H$ gates are invoked with the complexity $O(\log m)$. In step (2.2), it takes orales $U$ and $V$ with complexity $O[\log(m)+\max\{n_i\}]$. The complexity of performing oracle $V$ in step (2.3) is $O[$poly$\log(m\max\{n_i\})]$. In step (2.5), it takes $\max\{n_i\}$ times the quantum multiply-adder to obtain the inverse of the local density of $\bm{x^i}$, thus the complexity to prepare the state in Eq. (21) is $O[\max\{n_i\}\log(m)]$. The oracles $G$ and $W$ in steps (2.6)-(2.7) are performed with complexity $O[\log(m\max\{n_i\})]$. Similar to steps (2.1)-(2.3), the complexity of step (2.8) is $O[\log(m\max\{n_i\})+\max\{n_i\}]$. Step (2.9) repeats the operations of steps (2.4)-(2.5) to obtain Eq. (25) with complexity $O[\max\{n_i\}\log(m)]$. The total complexity is $O[\max\{n_i\}\log(m\max\{n_i\})]$. The complexity of steps (2.1) to (2.9) are illustrated in Table 2.
\begin{table}[htbp]
\centering
\label{TAB:table1}
\caption{The time complexity of steps (2.1) to (2.7)}
\setlength{\tabcolsep}{5mm}{
\begin{tabular}{ccccccc|cc}
\hline\hline
Steps\par&Complexity\\ \hline
(2.1)&$O(\log m)$\\
(2.2)&$O[\log(m)+\max\{n_i\}]$\\
(2.3)&$O[$poly$\log(m\max\{n_i\})]$\\
(2.4)-(2.5)&$O[\max\{n_i\}\log(m)]$\\
(2.6)-(2.7)&$O[\log(m\max\{n_i\})]$\\
(2.8)&$O[\log(m\max\{n_i\})+\max\{n_i\}]$\\
(2.9)&$O[\max\{n_i\}\log(m)]$\\
Total complexity&$O[\max\{n_i\}\log(m\max\{n_i\})]$\\
 \hline\hline
\end{tabular}}
\label{Table1}
\end{table}
\subsection{Quantum process of obtaining local outlier factors}
\label{sec3.3}
We have obtained the local reachability density of each data point, thus we can compute the local outlier factor of each data point in parallel using the amplitude estimate.
\subsubsection{Step details}
We calculate local outlier factors in parallel through the amplitude estimation and perform Grover's algorithm to search for abnormal data points satisfying $LOF_k(\bm{x^i})\geq\delta$. The whole procedure is depicted as follows.

(3.1) Perform the quantum multiply-adder on Eq. (25) to get
\begin{align}
&\frac{1}{\sqrt{m}}\sum_{i=1}^m|i\rangle\frac{1}{\sqrt{n_i}}\sum_{j=1,\bm{x^t}\in N_{k}(\bm{x^i})}^{n_i}|j\rangle|t\rangle|\frac{\big[\overline{lrd}_{k}(\bm{x^i})\big]^{-1}}{\big[\overline{lrd}_{k}(\bm{x^t})\big]^{-1}}\rangle\nonumber\\
&:=\frac{1}{\sqrt{m}}\sum_{i=1}^m|i\rangle\frac{1}{\sqrt{n_i}}\sum_{j=1,\bm{x^t}\in N_{k}(\bm{x^i})}^{n_i}|j\rangle|t\rangle|\rho_i^t\rangle,
\end{align}
where $\rho_i^t=\frac{\big[\overline{lrd}_{k}(\bm{x^i})\big]^{-1}}{\big[\overline{lrd}_{k}(\bm{x^t})\big]^{-1}}=\frac{\frac{1}{\sqrt{n}C}[lrd_{k}(\bm{x^i})]^{-1}}{\frac{1}{\sqrt{n}C}[lrd_{k}(\bm{x^t})]^{-1}}=\frac{lrd_{k}(\bm{x^t})}{lrd_{k}(\bm{x^i})}$. 

(3.2) Append an ancillary qubit and perform controlled rotate from $|0\rangle$ to $\sqrt{\frac{\rho_i^t}{E}}|0\rangle+\sqrt{1-\frac{\rho_i^t}{E}}|1\rangle$ conditioned on $|\rho_i^t\rangle$, we can obtain
\begin{equation}
\frac{1}{\sqrt{m}}\sum_{i=1}^m|i\rangle\frac{1}{\sqrt{n_i}}\sum_{j=1,\bm{x^t}\in N_{k}(\bm{x^i})}^{n_i}|j\rangle|t\rangle|\rho_i^t\rangle(\sqrt{\frac{\rho_i^t}{E}}|0\rangle+\sqrt{1-\frac{\rho_i^t}{E}}|1\rangle),
\end{equation}
where $E=\max\{\frac{lrd_{k}(\bm{x^t})}{lrd_{k}(\bm{x^i})}\}$. Similar to step (1.4), let $\sin^2(\alpha_i^t)=\frac{1}{n_i}\sum_{\bm{x^t}\in N_{k}(\bm{x^i})}\frac{\rho_i^t}{E}$ indicates the probability that $|0\rangle$ is measured, which can be obtained by amplitude estimation.

(3.3) Append an ancilla register, execute the quantum amplitude estimation to obtain and store the amplitude of $|0\rangle$ into the register, i.e.,
\begin{equation}
\frac{1}{\sqrt{m}}\sum_{i=1}^m|i\rangle|E\cdot\sin^2(\alpha_i^t)\rangle=\frac{1}{\sqrt{m}}\sum_{i=1}^m|i\rangle|LOF_{k}(\bm{x^i})\rangle.
\end{equation}

(3.4) Perform Grover's algorithm to search all indices $i$ of abnormal data points that satisfy $LOF_k(\bm{x^i})\geq\delta$.

\subsubsection{Complexity analysis}
The complexity of this step is $O(\frac{\sqrt{mT}\max\{n_i\}\log(m\max\{n_i\})}{\varepsilon_3})$, which is mainly derived from the amplitude estimation and the Grover's algorithm, where $T$ represents the number of abnormal data points and $\varepsilon_3$ represents the error of step (3.3). The specific analysis is as follows.

In step (3.3), the amplitude estimation block needs $O(1/\varepsilon_3)$ applications of the operators of preparing Eq. (27) to achieve error $\varepsilon_3$, where the complexity of performing the operators to obtain Eq. (27) is $O[\max\{n_i\}\cdot \log(m\max\{n_i\})]$. Therefore, the complexity of step (3.3) is $O(\frac{\max\{n_i\}\log(m\max\{n_i\})}{\varepsilon_3})$. In step (3.4), the complexity of executing Grover's algorithm to obtain all abnormal data points is
$O(\sqrt{mT}\cdot\frac{\max\{n_i\}\log(m\max\{n_i\})}{\varepsilon_2})$.

The error of the amplitude estimation performed in step (3.3) is
\begin{align}
&\big|\hat{LOF_k}(\bm{x^i})-\frac{1}{n_i}\sum_{\bm{x^t}\in N_{k}(\bm{x^i})}\frac{\hat{\overline{lrd}}_k(\bm{x^t})}{\hat{\overline{lrd}}_k(\bm{x^i})}\big|=E|\sin^2(\alpha_i^t)-\sin^2(\hat{\alpha}_i^t)|\nonumber\\
&\le E|\hat{\alpha}^t_i-\alpha^t_i|\le E\cdot\varepsilon_3.
\end{align}
Now, we analyze the errors of $LOF_k(\bm{x^i})$ as follow:
\begin{widetext}
	\begin{align}
&|\hat{LOF}_k(\bm{x^i})-LOF_k(\bm{x^i})|=\big|\hat{LOF_k}(i)-\frac{1}{n_i}\sum_{\bm{x^t}\in N_{k}(\bm{x^i})}\frac{lrd_k(\bm{x^t})}{lrd_k(\bm{x^i})}\big|\nonumber\\
&=|\hat{LOF}_k(\bm{x^i})-\frac{1}{n_i}\sum_{\bm{x^t}\in N_{k}(\bm{x^i})}\frac{\hat{\overline{lrd}}_k(\bm{x^t})}{\hat{\overline{lrd}}_k(\bm{x^i})}+\frac{1}{n_i}\sum_{\bm{x^t}\in N_{k}(\bm{x^i})}\big(\frac{\hat{\overline{lrd}}_k(\bm{x^t})}{\hat{\overline{lrd}}_k(\bm{x^i})}-\frac{lrd_k(\bm{x^t})}{lrd_k(\bm{x^i})}\big)|\nonumber\\
&\le \big|\hat{LOF}_k(\bm{x^i})-\frac{1}{n_i}\sum_{\bm{x^t}\in N_{k}(\bm{x^i})}\frac{\hat{\overline{lrd}}_k(\bm{x^t})}{\hat{\overline{lrd}}_k(\bm{x^i})}\big|+\big|\frac{1}{n_i}\sum_{\bm{x^t}\in N_{k}(\bm{x^i})}\big(\frac{\hat{\overline{lrd}}_k(\bm{x^t})}{\hat{\overline{lrd}}_k(\bm{x^i})}-\frac{\overline{lrd}_k(\bm{x^t})}{\overline{lrd}_k(\bm{x^i})}\big)\big|\nonumber\\
&\le E\cdot\varepsilon_3+\big|\frac{1}{n_i}\sum_{\bm{x^t}\in N_{k}(\bm{x^i})}\bigg(\frac{\frac{1}{n_i}\sum_{\bm{x^t}\in N_{k}(\bm{x^i})}\hat{\overline{reach\mbox{-}dist}}_{k}(\bm{x^i,x^t})}{\frac{1}{n_t}\sum_{\bm{x^l}\in N_{k}(\bm{x^t})}\hat{\overline{reach\mbox{-}dist}}_{k}(\bm{x^t,x^l})}-\frac{\frac{1}{n_i}\sum_{\bm{x^t}\in N_{k}(\bm{x^i})}\overline{reach\mbox{-}dist}_{k}(\bm{x^i,x^t})}{\frac{1}{n_t}\sum_{\bm{x^l}\in N_{k}(\bm{x^t})}\overline{reach\mbox{-}dist}_{k}(\bm{x^t,x^l})}\bigg)\big|\nonumber\\
&\le E\cdot\varepsilon_3+\big|\frac{1}{n_i}\sum_{\bm{x^t}\in N_{k}(\bm{x^i})}\bigg(\frac{\frac{1}{n_i}\sum_{\bm{x^t}\in N_{k}(\bm{x^i})}\hat{\overline{d}}(\bm{x^i,x^t})}{\frac{1}{n_t}\sum_{\bm{x^l}\in N_{k}(\bm{x^t})}\hat{\overline{d}}(\bm{x^t,x^l})}-\frac{\frac{1}{n_i}\sum_{\bm{x^t}\in N_{k}(\bm{x^i})}\overline{d}(\bm{x^i,x^t})}{\frac{1}{n_t}\sum_{\bm{x^l}\in N_{k}(\bm{x^t})}\overline{d}(\bm{x^t,x^l})}\bigg)\big|\nonumber\\
&\le E\cdot\varepsilon_3+\big|\frac{1}{n_i}\sum_{\bm{x^t}\in N_{k}(\bm{x^i})}\big(\frac{\frac{1}{n_i}\sum_{\bm{x^t}\in N_{k}(\bm{x^i})}\frac{1}{n_t}\sum_{\bm{x^l}\in N_{k}(\bm{x^t})}\big(\hat{\overline{d}}(\bm{x^i,x^t})\overline{d}(\bm{x^t,x^l})-\overline{d}(\bm{x^i,x^t})\hat{\overline{d}}(\bm{x^t,x^l})\big)}{(\frac{1}{n_t}\sum_{\bm{x^l}\in N_{k}(\bm{x^t})}\overline{d}(\bm{x^t,x^l}))\cdot(\frac{1}{n_t}\sum_{\bm{x^l}\in N_{k}(\bm{x^t})}\hat{\overline{d}}(\bm{x^t,x^l}))}\big)\big|\nonumber
\end{align}
\end{widetext}
\begin{widetext}
	\begin{align}
&\le E\cdot\varepsilon_3+\big|\frac{1}{n_i}\sum_{\bm{x^t}\in N_{k}(\bm{x^i})}\frac{\overline{d}(\bm{x^t,x^l})\varepsilon_1+\overline{d}(\bm{x^i,x^t})\varepsilon_1}{(\frac{1}{n_t}\sum_{\bm{x^l}\in N_{k}(\bm{x^t})}\overline{d}(\bm{x^t,x^l}))\cdot(\frac{1}{n_t}\sum_{\bm{x^l}\in N_{k}(\bm{x^t})}\hat{\overline{d}}(\bm{x^t,x^l}))}\big|\nonumber\\
&\le E\cdot\varepsilon_3+\frac{2\varepsilon_1}{(\frac{1}{n_t}\sum_{\bm{x^l}\in N_{k}(\bm{x^t})}\overline{d}(\bm{x^t,x^l}))\cdot(\frac{1}{n_t}\sum_{\bm{x^l}\in N_{k}(\bm{x^t})}\hat{\overline{d}}(\bm{x^t,x^l}))}.
\end{align}
\end{widetext}
We assume that at least half of the values of $\{\overline{d}(\bm{x^t,x^l})\}_{l=1}^{n_t}$ are greater than a constant $\sqrt{P}$, i.e.,
\begin{equation}
\sum_{\bm{x^l}\in N_{k}(\bm{x^t})}\overline{d}(\bm{x^t,x^l})\ge\frac{n_t}{2}\sqrt{P}, \sum_{\bm{x^l}\in N_{k}(\bm{x^t})}\hat{\overline{d}}(\bm{x^t,x^l})\ge\frac{n_t}{2}\sqrt{P}.
\end{equation}
The second term of Eq. (30) is as follows:
\begin{align}
&\frac{2\varepsilon_1}{(\frac{1}{n_t}\sum_{\bm{x^l}\in N_{k}(\bm{x^t})}\overline{d}(\bm{x^t,x^l}))\cdot(\frac{1}{n_t}\sum_{\bm{x^l}\in N_{k}(\bm{x^t})}\hat{\overline{d}}(\bm{x^t,x^l}))}\nonumber\\
&\le\frac{2\varepsilon_1}{\frac{1}{2}\sqrt{P}\cdot\frac{1}{2}\sqrt{P}}=\frac{8\varepsilon_1}{P}.
\end{align}
Therefore, we get $LOF_k(\bm{x^i})$ with error $E\varepsilon_2+\frac{4\varepsilon_1}{P}$.
\subsection{The total complexity}
The quantum algorithm can be divided into three steps and the complexity of each step can be seen in Table 3. Putting it all together, the complexity of the quantum LOF algorithm is $O[\max\{n_i\}\cdot \frac{m^{\frac{3}{2}}\log(mn)}{\varepsilon_1}+\sqrt{mT}\cdot\frac{\max\{n_i\}\log(m\max\{n_i\})}{\varepsilon_3}]$.

\begin{table}[htbp]
\centering
\label{TAB:table1}
\caption{The time complexity of the three steps of the quantum LOF algorithm}
\setlength{\tabcolsep}{8mm}{
\begin{tabular}{ccccccc|cc}
\hline\hline
Step\par&Complexity\\ \hline
Step 1&$O[\frac{\max\{n_i\}\cdot m^{\frac{3}{2}}\log(mn)}{\varepsilon_1\varepsilon_2}]$\\
Step 2&$O[\max\{n_i\}\log(m\max\{n_i\})]$\\
Step 3&$O[\sqrt{mT}\cdot\frac{\max\{n_i\}\log(m\max\{n_i\})}{\varepsilon_3}]$\\
 \hline\hline
\end{tabular}}
\label{Table1}
\end{table}

If $\max\{n_i\}=O(k), C,E,P=O(1), \varepsilon_1=\frac{P\varepsilon}{16}, \varepsilon_2=\varepsilon, \varepsilon_3=\frac{\varepsilon}{2E}$ and in general $T\ll m$, we can get $|\hat{LOF}_k(\bm{x^i})-LOF_k(\bm{x^i})|\le\varepsilon$. The overall runtime will be $O[k\cdot m^{\frac{3}{2}}\log(mn)/\varepsilon^2]$. It is shown that our quantum algorithm achieves polynomial speedup on $m$ and exponential speedup on $n$ compared to its classical counterpart.
\section{Conclusion}
\label{sec4}

In the present study, we propose a quantum LOF algorithm. It is shown that our quantum algorithm can achieve exponential speedup on the dimension of data points $n$ and polynomial speedup on the number $m$ of data points compared to its classical counterpart.

In the step 2, we proposed an efficient method to compute the local reachability density of each data point in parallel, which can be revisited as a subroutine of other quantum clustering algorithms and quantum dimensionality reduction algorithms. In step (2.3), the reason we can calculate the mean of reachability distance between each data point and its k-distance neighbors by the quantum multiply-adder is that we have managed to encode the distance information into the computational basis. This idea could also be applied to solve other machine learning problems, such as such as density estimation and feature learning. We hope the techniques used in our algorithm can inspire more anomaly detection algorithms to get a quantum advantage, especially unsupervised anomaly detection.

\section*{Acknowledgements}
\label{}
We thank HaiLing Liu, GuangHui Li and Di Zhang for useful discussions on the subject. This work is supported by the National Natural Science Foundation of China (Grants No.61976024, No.61972048, No.62272056) and supported by the 111 Project B21049.\\
\\

\end{document}